\documentclass[sigconf]{acmart}

\usepackage{fancyvrb}
\usepackage{url}
\usepackage{algorithmic}
\usepackage{textcomp}
\usepackage{xspace}
\usepackage{subcaption}
\usepackage{enumitem}
\usepackage{listings} 
\usepackage{multirow} 
\usepackage{listings}

\newcommand{\rqa}{$RQ_1$}
\newcommand{\rqb}{$RQ_2$}
\newcommand{\rqc}{$RQ_3$}

\newcommand{\rqaa}{How prevalent are readability-related commits generated by AI agents?}
\newcommand{\rqbb}{What issues do agent-generated readability commits address?}
\newcommand{\rqcc}{How does code readability differ before and after AI agent-generated readability commits?}

\newcommand{\rqA}{\rqa: \rqaa}
\newcommand{\rqB}{\rqb: \rqbb}
\newcommand{\rqC}{\rqc: \rqcc}

\newcommand{\CatComplex}{{\small\sc Complex, Long, or Inadequate Logic}\xspace}

\newcommand{\CatDoc}{{\small\sc Incomplete or Inadequate Code Documentation}\xspace}
\newcommand{\CatIdentifier}{{\small\sc Bad Identifier}\xspace}

\newcommand{\CatString}{{\small\sc Wrong, Missing, or Inadequate String Expression or Literal}\xspace}
\newcommand{\CatConstant}{{\small\sc Missing Constant Usage}\xspace}

\newcommand{\readsmells}{\textit{readability smells}\xspace}
\newcommand{\loc}{\textit{Lines of Code}\xspace}
\newcommand{\lloc}{\textit{Logical Lines of Code}\xspace}
\newcommand{\sloc}{\textit{Source Lines of Code}\xspace}
\newcommand{\mi}{\textit{Maintainability Index}\xspace}
\newcommand{\cc}{\textit{Cyclomatic Complexity}\xspace}
\newcommand{\hv}{\textit{Halstead Volume}\xspace}
\newcommand{\hd}{\textit{Halstead Difficulty}\xspace}
\newcommand{\he}{\textit{Halstead Effort}\xspace}

\newcommand{\FullTotalCommits}{1,382,314\xspace}
\newcommand{\readabilityCommits}{4,115\xspace}
\newcommand{\readabilityPercentage}{0.3}
\newcommand{\pythonReadabilityCommits}{577\xspace}
\newcommand{\avgLabels}{1.3\xspace}

\newcommand{\keywordReadability}{2,061\xspace}
\newcommand{\keywordClarity}{2,003\xspace}
\newcommand{\keywordReadable}{254\xspace}
\newcommand{\keywordLegibility}{16\xspace}
\newcommand{\keywordEasierToRead}{7\xspace}
\newcommand{\keywordUnderstandabe}{3\xspace}
\newcommand{\keywordUnderstandability}{1\xspace}
\newcommand{\keywordComprehensible}{0\xspace}

\newcommand{\pctReadability}{50.1}
\newcommand{\pctClarity}{48.7}
\newcommand{\pctReadable}{6.2}
\newcommand{\pctLegibility}{0.4}
\newcommand{\pctEasierToRead}{0.2}
\newcommand{\pctUnderstandable}{0.1}
\newcommand{\pctUnderstandability}{0.0}
\newcommand{\pctComprehensible}{0.0}

\newcommand{\sampledCommits}{231\xspace}

\newcommand{\labelAgreement}{94.7\xspace}
\newcommand{\cohensKappa}{0.80\xspace}
\newcommand{\disagreementLabels}{98\xspace}
\newcommand{\totalLabels}{1,832\xspace}

\newcommand{\catComplexPercent}{42.4}

\newcommand{\catDocPercent}{24.2}

\newcommand{\catIdentifierPercent}{14.3}

\newcommand{\catFormattingPercent}{14.3}

\newcommand{\catLoggingPercent}{12.6}

\newcommand{\catUnnecessaryPercent}{10.0}

\newcommand{\catStringPercent}{6.1}

\newcommand{\catConstantPercent}{2.6}

\newcommand{\catHumanComplexPercent}{18.2}
\newcommand{\catHumanDocPercent}{22.3}
\newcommand{\catHumanIdentifierPercent}{20.3}
\newcommand{\catHumanFormattingPercent}{10.9}
\newcommand{\catHumanLoggingPercent}{4.7}
\newcommand{\catHumanUnnecessaryPercent}{13.2}
\newcommand{\catHumanStringPercent}{8.1}
\newcommand{\catHumanConstantPercent}{2.3}

\newcommand{\validCommits}{403\xspace}

\newcommand{\miMeanDiff}{-3.25\xspace}
\newcommand{\miMedianDiff}{-0.12\xspace}
\newcommand{\miEffectSize}{-0.35\xspace}
\newcommand{\miPValue}{<0.001\xspace}

\newcommand{\ccMeanDiff}{3.13\xspace}
\newcommand{\ccMedianDiff}{0.00\xspace}
\newcommand{\ccEffectSize}{0.63\xspace}
\newcommand{\ccPValue}{<0.001\xspace}

\newcommand{\locMeanDiff}{27.61\xspace}
\newcommand{\locMedianDiff}{6.00\xspace}
\newcommand{\locEffectSize}{0.64\xspace}
\newcommand{\locPValue}{<0.001\xspace}

\newcommand{\llocMeanDiff}{12.18\xspace}
\newcommand{\llocMedianDiff}{1.00\xspace}
\newcommand{\llocEffectSize}{0.54\xspace}
\newcommand{\llocPValue}{<0.001\xspace}

\newcommand{\slocMeanDiff}{18.47\xspace}
\newcommand{\slocMedianDiff}{2.60\xspace}
\newcommand{\slocEffectSize}{0.60\xspace}
\newcommand{\slocPValue}{<0.001\xspace}

\newcommand{\cmtMeanDiff}{1.67\xspace}
\newcommand{\cmtMedianDiff}{0.00\xspace}
\newcommand{\cmtEffectSize}{0.30\xspace}
\newcommand{\cmtPValue}{<0.001\xspace}

\newcommand{\slcMeanDiff}{0.73\xspace}
\newcommand{\slcMedianDiff}{0.00\xspace}
\newcommand{\slcEffectSize}{0.17\xspace}
\newcommand{\slcPValue}{0.022\xspace}

\newcommand{\mlcMeanDiff}{2.10\xspace}
\newcommand{\mlcMedianDiff}{0.00\xspace}
\newcommand{\mlcEffectSize}{0.63\xspace}
\newcommand{\mlcPValue}{<0.001\xspace}

\newcommand{\hvMeanDiff}{43.60\xspace}
\newcommand{\hvMedianDiff}{0.00\xspace}
\newcommand{\hvEffectSize}{0.50\xspace}
\newcommand{\hvPValue}{<0.001\xspace}

\newcommand{\hdMeanDiff}{0.31\xspace}
\newcommand{\hdMedianDiff}{0.00\xspace}
\newcommand{\hdEffectSize}{0.35\xspace}
\newcommand{\hdPValue}{<0.001\xspace}

\newcommand{\heMeanDiff}{321.49\xspace}
\newcommand{\heMedianDiff}{0.00\xspace}
\newcommand{\heEffectSize}{0.42\xspace}
\newcommand{\hePValue}{<0.001\xspace}

\newcommand{\miWorsenPercent}{56.1}

\newcommand{\ccImprovePercent}{9.7}
\newcommand{\ccWorsenPercent}{42.7}

\newcommand{\locWorsenPercent}{71.5}

\definecolor{darkgreen}{rgb}{0, 0.5, 0} 
\definecolor{whitesmoke}{rgb}{0.99, 0.99, 0.99} 

\def\Underline{\setbox0\hbox\bgroup\let\\\endUnderline}
\def\endUnderline{\vphantom{y}\egroup\smash{\underline{\box0}}\\}
\def\|{\verb|}

\newcommand{\ie}{\textit{i.e.,}\xspace}

\newcommand{\etal}{\xspace\textit{et al.}\xspace}

\newcommand{\motivation}{\noindent\textbf{Motivation. }}
\newcommand{\approach}{\smallskip\noindent\textbf{Approach. }}
\newcommand{\results}{\smallskip\noindent\textbf{Results. }}

\newcounter{findings_no}

\usepackage{listings}
\definecolor{backcolour}{rgb}{0.95,0.95,0.92}
\lstdefinelanguage{diff}{
  numberstyle={\tiny},
  morecomment=**[f][\color{red}]{-},         
  morecomment=**[f][\color{darkgreen}]{+},       
  moredelim=**[is][\bfseries\color{red}]{<@}{@>},
  moredelim=**[is][\bfseries\color{darkgreen}]{<+}{+>},
  basicstyle={\ttfamily \tiny}
}
\definecolor{backcolour}{rgb}{0.95,0.95,0.92}
\lstdefinelanguage{commit}{ 
  breakindent = 0pt,
  numbers=none,
  backgroundcolor=\color{white},
  frame=single,
  xleftmargin=3.5em,
  numbersep=0em,
  xrightmargin=1.5em,
}

\makeatletter
\newsavebox{\dbox@box}
\newlength{\dbox@innerwd}
\newenvironment{dbox}{%
  \par\medskip\noindent
  \setlength{\fboxsep}{4pt}%
  \setlength{\fboxrule}{1pt}%
  \setlength{\dbox@innerwd}{\dimexpr\linewidth-2\fboxsep-2\fboxrule\relax}%
  \begin{lrbox}{\dbox@box}%
  \begin{minipage}{\dbox@innerwd}%
  \setlength{\parindent}{0pt}%
}{%
  \end{minipage}%
  \end{lrbox}%
  \noindent\fcolorbox{black}{gray!15}{\usebox{\dbox@box}}%
  \par\medskip
}
\makeatother

\begin{document}




\AtBeginDocument{%
  \providecommand\BibTeX{{%
    \normalfont B\kern-0.5em{\scshape i\kern-0.25em b}\kern-0.8em\TeX}}}



%
%

\copyrightyear{2026}
\acmYear{2026}
\setcopyright{cc}
\setcctype{by}
\acmConference[MSR ’26]{23rd International Conference on Mining Software Repositories}{April 13--14, 2026}{Rio de Janeiro, Brazil}
\acmBooktitle{23rd International Conference on Mining Software Repositories (MSR ’26), April 13--14, 2026, Rio de Janeiro, Brazil}
\acmPrice{}
\acmDOI{10.1145/3793302.3793590}
\acmISBN{979-8-4007-2474-9/2026/04}

\title{Do AI Agents Really Improve Code Readability?}






\author{
  Kyogo Horikawa\textsuperscript{$\dagger$},
  Kosei Horikawa\textsuperscript{$\S$},
  Yutaro Kashiwa\textsuperscript{$\S$},
  Hidetake Uwano\textsuperscript{$\dagger$},
  Hajimu Iida\textsuperscript{$\S$}
}

\affiliation{%
  \institution{\textsuperscript{$\dagger$}National Institute of Technology, Nara College, Japan}
  \country{}
}
\affiliation{%
  \institution{\textsuperscript{$\S$}Nara Institute of Science and Technology, Japan}
  \country{}
}

\email{AI1221@nara.kosen-ac.jp, uwano@info.nara-k.ac.jp}
\email{horikawa.kosei.hk1@naist.ac.jp, yutaro.kashiwa@is.naist.jp, iida@itc.naist.jp}

\renewcommand{\shortauthors}{Horikawa, et al.}

\begin{abstract}
Code readability is fundamental to software quality and maintainability. Poor readability extends development time, increases bug-inducing risks, and contributes to technical debt. With the rapid advancement of Large Language Models, AI agent-based approaches have emerged as a promising paradigm for automated refactoring, capable of decomposing complex tasks through autonomous planning and execution. While prior studies have examined refactoring by AI agents, these analyses cover all forms of refactoring, including performance optimization and structural improvement. As a result, the extent to which AI agent-based refactoring specifically improves code readability remains unclear.

This study investigates the impact of AI agent-based refactoring on code readability. We extracted commits containing readability-related keywords from the AIDev dataset and analyzed changes in readability metrics before and after each commit, covering \validCommits commits evaluated using multiple quantitative metrics. Our results indicate that AI agents primarily target logic complexity (\catComplexPercent\%) and documentation improvements (\catDocPercent\%) rather than surface-level aspects like naming conventions or formatting. However, contrary to expectations, readability-focused commits often degraded traditional quality metrics: the \mi decreased in \miWorsenPercent\% of commits, while \cc increased in \ccWorsenPercent\%.

\end{abstract}



\maketitle

\section{Introduction}\label{sec:introduction}

Code readability is a critical factor influencing software maintainability and sustainability~\cite{DBLP:journals/tse/OliveiraSOMCM25}.
Poor readability leads to technical debt~\cite{DBLP:journals/corr/abs-2510-23068, DBLP:conf/msr/ChowdhuryKA25} and increased communication costs~\cite{DBLP:journals/tse/BuseW10, DBLP:conf/msr/PosnettHD11, DBLP:journals/tse/ScalabrinoBVLPO21}. 
To address this, software engineering research has proposed various approaches, including automated refactoring tools~\cite{DBLP:journals/pacmse/GaoHYX25,DBLP:journals/jss/ShahidiAN22, DBLP:conf/issta/LiuWWXWLJ23}, quantitative metrics~\cite{DBLP:conf/issta/BuseW08, DBLP:journals/smr/ScalabrinoLOP18}, and machine learning-based prediction models~\cite{DBLP:conf/iwpc/ScalabrinoVPO16, DBLP:conf/iwpc/RoyFLA20}.

While these traditional approaches have reduced developer burden, the recent emergence of Large Language Models (LLMs) has elevated automated refactoring to a new level of practicality.
Particularly since the emergence of conversational LLMs such as ChatGPT, numerous practical applications have been reported~\cite{DBLP:conf/ease/WatanabeK0HYI24, DBLP:conf/fose-ws/FanGHLSYZ23}. In this workflow, developers provide refactoring instructions through prompts, and LLMs immediately present code improvement suggestions, achieving significant efficiency gains compared to conventional methods. Cases of LLM-assisted refactoring are actively shared in Stack Overflow and GitHub discussions~\cite{DBLP:journals/corr/abs-2411-02320, DBLP:conf/msr/TufanoMPDPB24}, indicating that this technology is becoming increasingly adopted in development practice.

More recently, there has been a shift from single prompt-based approaches to AI agent-based approaches. These approaches enable the decomposition of complex tasks into multiple subtasks, allowing for more sophisticated refactoring through autonomous cycles of planning, execution, and verification~\cite{DBLP:journals/corr/abs-2509-14745}. Existing research has analyzed what types of refactoring AI agents perform and to what extent~\cite{DBLP:journals/corr/abs-2511-04824}. However, these studies examine refactoring activities in general, which include not only readability-focused refactoring but also refactoring aimed at performance optimization~\cite{DBLP:journals/tosem/CortellessaPPFJTH25}, structural improvement~\cite{DBLP:conf/wcre/BoisDV04}, and other objectives~\cite{DBLP:conf/icsm/IammarinoZAP19}. As a result, the specific impact of AI agent-based refactoring on code readability has not been sufficiently clarified. This represents an important research gap, as understanding the effectiveness of agent-based approaches for readability improvement is essential for developers who seek to leverage these tools for this particular purpose.

Therefore, this study aims to clarify how much AI agent-based refactoring contributes to code readability. Specifically, we extracted commits containing keywords related to readability from the AIDev dataset~\cite{DBLP:journals/corr/abs-2507-15003} and investigated the impact of these commits on readability. We identified \validCommits commits and calculated three readability metrics for the code before and after each commit, allowing a thorough analysis of the effects of agent-based refactoring. Our findings show that readability-oriented commits do not necessarily lead to improvements in structural code quality and are often accompanied by quality degradation.


\begin{figure}[h]
  \centering
  \includegraphics[width=0.95\linewidth]{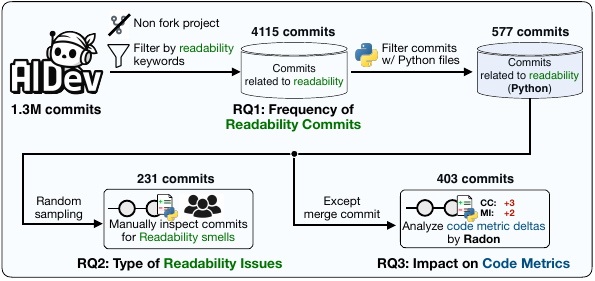}
  \vspace{-0.2cm}
  \caption{Overview of the study design}
  \Description{A diagram showing the overview of the study design.}
  \label{fig:overview}
  \vspace{-0.5cm}
\end{figure}

\section{Related Work}\label{sec:relatedwork}

\textbf{Code Readability and Assessment. }Code readability is central to maintainability, prompting extensive research into its quantification.
Buse and Weimer~\cite{DBLP:journals/tse/BuseW10} and Scalabrino\etal~\cite{DBLP:journals/tse/ScalabrinoBVLPO21} proposed models to predict readability based on code features. Furthermore, empirical studies by Dantas\etal~\cite{DBLP:conf/icsm/DantasRM23} and Oliveira\etal~\cite{DBLP:journals/tse/OliveiraSOMCM25} classified how human developers improve readability in practice, identifying categories such as identifier renaming and comment additions. More recently, Takerngsaksiri\etal~\cite{DBLP:conf/icsm/TakerngsaksiriTFPCW25}
compared LLM-generated code with human-written code, finding comparable readability in enterprise settings. However, these studies focused on either human-driven improvements or code generation from scratch; whether AI agents can effectively improve the readability of existing code through refactoring has not been examined.

\noindent\textbf{Refactoring by LLMs and Agents. }The emergence of LLMs has advanced automated refactoring. While Cordeiro\etal~\cite{DBLP:journals/corr/abs-2411-02320} examined the refactoring capabilities of LLMs, their focus was on prompt-based approaches. More recently, autonomous coding agents have emerged~\cite{DBLP:journals/corr/abs-2507-15003, DBLP:journals/corr/abs-2509-14745}. The closest work to ours is Horikawa\etal~\cite{DBLP:journals/corr/abs-2511-04824}, who analyzed the impact of agent-driven refactoring on various code metrics. They did not observe any significant improvement in readability. 
 However, they collected commits with commit messages containing ``refactor''. This keyword is broad, and refactoring involves many operations including performance optimization~\cite{DBLP:journals/tosem/TrainiPTLSBLOC22}, code restructuring~\cite{DBLP:conf/wcre/MooijKKS20}, security enhancement~\cite{DBLP:journals/sncs/EdwardNE24, ALMOGAHED2022108}, and modularity improvement~\cite{DBLP:journals/sigsoft/YuR08}, beyond just readability. In this study, we focus on readability-improving commits made by agents and examine the impact on readability metrics established by Oliveira\etal~\cite{DBLP:journals/tse/OliveiraSOMCM25}.






\section{Data Collection}\label{sec:studydesign}
In this section, we describe the data mining and data filtering as outlined in \autoref{fig:overview}. 

\subsection{Data Mining}\label{subsec:datamining}
To study the relationship between code readability and AI Agents, we start with the AIDev dataset~\cite{DBLP:journals/corr/abs-2507-15003}.
This dataset contains 932,791 pull requests (PRs) created by five coding agents from over 61,000 repositories. However, the AIDev dataset only includes commit data for repositories with more than 100 GitHub stars. To gather a more comprehensive set of data points, we leverage the GitHub REST API\footnote{\url{https://docs.github.com/en/rest}} to mine all commits based on the provided metadata. In total, we collect \FullTotalCommits agentic commits.

\subsection{Identifying Commits improving Readability}
We then employ a keyword-based filtering approach on commit messages to extract commits related to readability improvement. 
Specifically, we performed case-insensitive filtering on commit messages using eight keywords derived from prior work [6]: ``readability'', ``readable'', ``understandability'', ``understandable'', ``clarity'', ``legibility'', ``easier to read'', and ``comprehensible''. This set of keywords was selected to broadly capture different expressions referring to improvements in code readability and understandability.
For RQ2 and RQ3, to ensure that the extracted commits correspond to source code readability improvements rather than documentation or configuration changes, we further filtered commits to those that include modifications to at least one Python source file.



\section{Research Questions}\label{sec:results}
\subsection*{\rqA}\label{sec:rqa}
\motivation 
Developers increasingly use AI agents for a wide range of software development tasks~\cite{DBLP:journals/corr/abs-2507-15003}. Prior work has observed that these agents frequently perform refactorings~\cite{DBLP:journals/corr/abs-2509-14745}, and Horikawa \etal\cite{DBLP:journals/corr/abs-2511-04824} examined how common such refactoring activities are. However, it is unclear to what extent these changes specifically target readability-improvements.


\approach
We flagged commits as readability-related if they contained at least one of these terms. We measure the frequency of readability-related commits and their percentage among all commits generated by AI agents (\ie all commits in the dataset).

\results
\autoref{tab:keyword_count} shows the prevalence of readability-related commits, which explicitly describe code comprehensibility and readability. We found that \readabilityCommits commits contained keywords related to readability, accounting for approximately 0.3\% of all commits.
On average, each selected commit contained \avgLabels readability-related keywords.
The most frequent term in commit messages generated by AI agents was ``readability'', appearing in \pctReadability\% of the commits.

These results indicate that, based on explicit commit messages, changes intended to improve readability account for a limited fraction of all AI-generated commits. Furthermore, AI agents tend to use standardized and general-purpose vocabulary when describing readability-related changes.


\begin{table}[t]
\small
\centering
\caption{Frequency of readability keywords}
\label{tab:keyword_count}
\begin{tabular}{l r}
\toprule
Keyword & Number of Commits (\%) \\ \midrule
readability         & \keywordReadability (\pctReadability\%) \\
clarity             & \keywordClarity (\pctClarity\%) \\
readable            & \keywordReadable (\pctReadable\%)  \\
legibility          & \keywordLegibility (\pctLegibility\%)   \\
easier to read      & \keywordEasierToRead (\pctEasierToRead\%)    \\
understandable       & \keywordUnderstandabe (\pctUnderstandable\%)    \\
understandability   & \keywordUnderstandability (\pctUnderstandability\%)    \\
comprehensible      & \keywordComprehensible (\pctComprehensible\%)    \\
\bottomrule
\end{tabular}
\end{table}

\


\begin{dbox}
\textbf{Answer to RQ1.} AI agents seldom focus on readability, with only 0.3\% of agent-created commits related to it.
\end{dbox}


\subsection*{\rqB}\label{sec:rqb}
\motivation Prior research has identified several \readsmells that impede code comprehension, including poor identifier naming, insufficient comments, and excessive logic complexity~\cite{DBLP:journals/tse/OliveiraSOMCM25}. However, these studies focused on human-written code, and whether the same patterns appear in code generated or refactored by AI Agents has not been examined. Identifying the strengths and weaknesses of AI Agents in addressing \readsmells would help developers better understand when to rely on these tools and where human review is still necessary.

\approach
From the \pythonReadabilityCommits readability-related commits modifying Python files, we randomly sampled \sampledCommits commits to ensure a 95\% confidence level with a 5\% margin of error. To ensure classification reliability, we employed a closed card sorting approach~\cite{spencer2009card}. Two authors, with 5 to 8 years of software development experience, independently inspected the commit messages and diff of the sampled commits. They mapped each commit to the eight \readsmells defined by Oliveira\etal\cite{DBLP:journals/tse/OliveiraSOMCM25}. Since a commit may address multiple readability aspects, multi-labeling was permitted.

Following independent annotation, we evaluated inter-rater reliability. The initial classification achieved \labelAgreement\% label-level agreement (Micro-averaged Cohen's Kappa: \cohensKappa), with disagreements occurring on \disagreementLabels of \totalLabels labels. For these conflicting cases, a third inspector reviewed the labels and recommended resolutions. These recommendations were discussed among all three authors until complete consensus was reached.

\begin{table}[t]
\centering
\caption{Addressed Readability Smells: Human vs. Agent}
\label{tab:categories}
\begin{tabular}{l r r}
\toprule
Category & Human~\cite{DBLP:journals/tse/OliveiraSOMCM25} & Agent \\
\midrule
Complex, Long, or Inadequate Logic              & \catHumanComplexPercent\%     & \catComplexPercent\% \\
Incomplete or Inadequate Code Doc.     & \catHumanDocPercent\%         & \catDocPercent\% \\
Bad Identifier                                  & \catHumanIdentifierPercent\%  & \catIdentifierPercent\% \\
Inconsistent or Disrupted Formatting            & \catHumanFormattingPercent\%  & \catFormattingPercent\% \\
Inadequate Logging and Monitoring               & \catHumanLoggingPercent\%     & \catLoggingPercent\% \\
Unnecessary Code                                & \catHumanUnnecessaryPercent\% & \catUnnecessaryPercent\% \\
Wrong, Missing, or Inadequate String & \catHumanStringPercent\%      & \catStringPercent\% \\
Missing Constant Usage                          & \catHumanConstantPercent\%    & \catConstantPercent\% \\
\bottomrule
\end{tabular}
\end{table}


\results \autoref{tab:categories} compares the prevalence of \readsmells addressed by AI agents versus human developers.
Our analysis reveals distinct differences in prioritization between the two groups.

First, regarding AI agents, the results show a heavily skewed distribution. The top two categories, \CatComplex and \CatDoc, are the most dominant, collectively accounting for the majority of improvements. In contrast, \CatConstant and \CatString were the least addressed categories, representing the bottom two with negligible percentages (\catConstantPercent\% and \catStringPercent\%, respectively).

Second, a comparison with the baseline shows that agents exhibit a much stronger bias than human developers. As shown in the table, human contributions are relatively evenly distributed across several categories, such as \CatDoc (\catHumanDocPercent\%), \CatIdentifier (\catHumanIdentifierPercent\%), and \CatComplex (\catHumanComplexPercent\%). AI agents, however, prioritize logic complexity and documentation over surface-level readability aspects.



To illustrate this prioritization, we examine a representative improvement in \CatComplex.
In Listing~\ref{code:diff_example_Complex}, a complex string operation involving nested formatting and exception handling was simplified using \texttt{json.dumps}. This change abstracts away manual character escaping, making the logic clearer and easier to understand. This example illustrates that AI agents prioritize improvements that enhance understandability and maintainability over minor stylistic adjustments.


\begin{lstlisting}[
    language=diff, 
    breaklines=true, 
    caption={[Example of Complex, Long, or Inadequate Logic]Example of Complex, Long, or Inadequate Logic\protect\footnotemark}, 
    label=code:diff_example_Complex, 
    aboveskip=0pt, 
    belowskip=10pt,
    escapechar=|
]
    yield rx.call_script(
-       f''try {{ signals.identify('<@{email.replace(chr(39), chr(92) + chr(39))}@>'); }} catch(e) {{ console.warn('Signals identify failed:', e); }}''
+       f''try {{ signals.identify('<+{json.dumps(email)}+>'); }} catch(e) {{ console.warn('Signals identify failed:', e); }}''
    )
\end{lstlisting}\footnotetext{\url{https://github.com/reflex-dev/reflex-web/commit/68afd5c9e677e626218e7dc4b2584ff5c1e61b3e}}

\begin{dbox}
  \textbf{Answer to RQ2.} AI agents primarily focus on improving logic complexity and explanatory clarity rather than surface-level readability aspects. 
\end{dbox}

\subsection*{\rqC}\label{sec:rqc}
\motivation Previous studies~\cite{DBLP:journals/tse/OliveiraSOMCM25} indicate that developers spend approximately 70\% of their time reading and understanding existing code, and they dedicate substantial maintenance effort to improving readability~\cite{DBLP:journals/tse/KimZN14, DBLP:journals/ese/PiantadosiFSSO20}. Moreover, prior work has shown that such activities can lead to measurable improvements in code readability over time ~\cite{DBLP:journals/ese/PiantadosiFSSO20}.
However, it is not clear to what extent AI agents improve readability in terms of quantitative metrics.


%
\approach
We extracted valid pre- and post-commit code states for \validCommits of 577 commits (excluding files with parsing errors).
Using the \texttt{radon} library~\cite{Radon}, we calculated the \mi, where higher values indicate better maintainability.
\mi is a composite metric that combines static code measures such as lines of code, cyclomatic complexity, and Halstead metrics.
We also calculated \cc, where lower values indicate simpler control structures.
To measure code size at different levels of granularity, we recorded \loc (total lines), \lloc, \sloc (excluding blank lines and comments), and \textit{Comment lines}.
Additionally, we considered Halstead metrics, namely \hv, \hd, and \he, to quantify different aspects of program complexity and cognitive effort.
We chose \texttt{radon} due to its widespread adoption in prior empirical software engineering research~\cite{DBLP:journals/corr/abs-2304-13187, DBLP:conf/nips/LuGRHSBCDJTLZSZ21, DBLP:conf/msr/IdialuMMAN24}.

We computed the change for each commit as $\Delta = \text{Metric}_{\text{after}} - \text{Metric}_{\text{before}}$.
To determine statistical significance, we employed the Wilcoxon signed-rank test~\cite{Wilcoxon1945} ($p < 0.01$).
We selected this non-parametric test because Q-Q plots exhibited S-shaped patterns, indicating that the paired differences did not follow a normal distribution.
Additionally, we quantified the magnitude of the impact using Cliff's delta~\cite{cliff1993dominance} as the effect size.

\results
\autoref{tab:rq3-mean-median} summarizes the statistics of the metric differences between the pre- and post-commit code states. Overall, the $p$-values for most of the metrics are below 0.01, confirming significant differences before and after readability-related commits. Note that the Wilcoxon signed-rank test excludes pairs with zero differences. Consequently, even when the median difference is zero, the test can detect significant differences if the non-zero changes are predominantly in one direction.

The median increase in \loc (\locMedianDiff) is more than double that of \sloc (\slocMedianDiff), indicating that AI-based changes affected comments and formatting beyond executable statements, though the non-zero median for \sloc confirms that modifications to executable logic were also present. For Halstead metrics, the median difference for \hv, \hd, and \he is zero, yet their mean values increased, implying that a specific subset of commits introduced substantial increases in code volume and cognitive effort.

As for effect sizes, \mi shows a medium effect size (\miEffectSize), indicating a moderate shift in maintainability, while structural and size-related metrics such as \cc (\ccEffectSize), \loc (\locEffectSize), and \sloc (\slocEffectSize) exhibit large effect sizes ($|r| \ge 0.5$), demonstrating strong and consistent increases in code size and complexity. Since a higher Maintainability Index is desirable while lower values are preferable for complexity and size metrics, these results suggest that AI-based readability-related commits actually degraded code quality from the perspective of traditional software metrics.

We also analyzed the proportions of improvement and deterioration across all commits. \mi deteriorated in \miWorsenPercent\% of the commits, indicating a tendency toward reduced maintainability. Similarly, \loc increased in \locWorsenPercent\% of the cases, while \cc stayed unchanged in nearly half of the commits, with only \ccImprovePercent\% showing improvement. This reinforces that readability-oriented commits often involve increasing code volume without simplifying control structures. 
\begin{table}
\centering
\caption{Metric changes in readability-related commits}
\label{tab:rq3-mean-median}
\begin{tabular}{lrrrr}
\toprule
Metric & Mean & Median & $p$-value & ES \\
\midrule

Lines of Code
  & \locMeanDiff & \locMedianDiff & \locPValue & \locEffectSize \\

Cyclomatic Complexity
  & \ccMeanDiff & \ccMedianDiff & \ccPValue & \ccEffectSize \\

Multi-line Comments
  & \mlcMeanDiff & \mlcMedianDiff & \mlcPValue & \mlcEffectSize \\
  
Source Lines of Code
  & \slocMeanDiff & \slocMedianDiff & \slocPValue & \slocEffectSize \\

Logical Lines of Code
  & \llocMeanDiff & \llocMedianDiff & \llocPValue & \llocEffectSize \\

Halstead Volume
  & \hvMeanDiff & \hvMedianDiff & \hvPValue & \hvEffectSize \\

Halstead Effort
  & \heMeanDiff & \heMedianDiff & \hePValue & \heEffectSize \\

Halstead Difficulty
  & \hdMeanDiff & \hdMedianDiff & \hdPValue & \hdEffectSize \\

Comment lines
  & \cmtMeanDiff & \cmtMedianDiff & \cmtPValue & \cmtEffectSize \\

Single-line Comments
  & \slcMeanDiff & \slcMedianDiff & \slcPValue & \slcEffectSize \\

Maintainability Index
  & \miMeanDiff & \miMedianDiff & \miPValue & \miEffectSize \\
\bottomrule
{\footnotesize *ES: Effect Size}
\end{tabular}

\end{table}
















\bigskip
\begin{dbox}
  \textbf{Answer to RQ3.} AI agent-generated readability commits do not necessarily improve structural code quality and often accompany increases in code volume and reduced maintainability.
\end{dbox}

\newpage
\section{Implication}

\noindent\textbf{\textit{Practitioners should critically review AI-proposed readability changes rather than accepting them automatically.}}
Our findings show that AI agents rarely prioritize readability (\readabilityPercentage\% in RQ1). Moreover, when they do, structural quality often degrades: \miWorsenPercent\% of commits worsened \mi and \ccWorsenPercent\% increased \cc (RQ3). This suggests that current agents lack a robust understanding of effective readability improvements.

\noindent\textbf{\textit{Researchers and developers should incorporate quality-checking mechanisms to prevent counterproductive refactoring.}}
Given this frequent structural degradation, tool developers should implement dedicated quality-checking agents. These agents can monitor metrics like Halstead complexity and code size (LOC) to reject refactorings that increase cognitive load. Future research should also enhance the verification phase, moving beyond functional correctness to include qualitative evaluations for better code organization.

\section{Threats to Validity}
\label{sec:threats_to_validity}

\noindent\textbf{Internal Validity:} Manual classification in RQ2 introduces subjectivity. We mitigated this by having three independent inspectors classify commits and resolve disagreements through discussion.

\noindent\textbf{Construct Validity:} Our analysis may be affected by tangled commits, where multiple unrelated changes are bundled together, meaning observed metric changes may reflect modifications beyond readability improvements. Additionally, metrics such as MI and CC may not fully capture all aspects of readability, such as clarity of intent and logical flow.

\noindent\textbf{External Validity:} Our strict filtering resulted in a small sample size (0.3\%), and our analysis focuses on Python projects, limiting generalizability to other languages. Future work should extend this study to additional languages and broader keyword sets.

\section{Conclusion}\label{sec:conclusion}
This study investigated the impact of AI agent-based refactoring on code readability in open-source software projects. Our analysis of \validCommits readability-related PRs shows that AI agents primarily focus on reducing logic complexity and improving documentation, with less attention to surface-level aspects such as naming conventions and formatting. However, these changes do not necessarily improve structural quality metrics and are often accompanied by decreased maintainability and increased complexity.

Future research should focus on developing readability-focused agents that incorporate qualitative evaluations and consider long-term maintainability, drawing on human developers' improvement practices. Such agents would go beyond surface-level changes, such as renaming identifiers or adding comments, and instead prioritize modifications that fundamentally enhance code comprehensibility through better organization at the functional level.

\noindent{\bf Replication Package:} The data and scripts used in our work is publicly available in the replication package~\cite{ReplicationPackage}.

\begin{acks}
This work was supported by JSPS KAKENHI (JP24K02921, JP25K21359) and JST PRESTO (JPMJPR22P3), ASPIRE (JPMJAP2415), and AIP Accelerated Program (JPMJCR25U7).

\end{acks}

\balance
\bibliographystyle{ACM-Reference-Format}
\bibliography{references}


\end{document}